\documentclass[%
 aip,
 cha,
 amsmath,amssymb,
 reprint,%
]{revtex4-1}

\usepackage{graphicx}
\usepackage{dcolumn}
\usepackage{bm}

\usepackage[utf8]{inputenc}
\usepackage[T1]{fontenc}
\usepackage{mathptmx}
\usepackage{etoolbox}

\usepackage[final]{hyperref} 
\hypersetup{
	colorlinks=true,       
	linkcolor=blue,        
	citecolor=blue,        
	filecolor=magenta,     
	urlcolor=blue         
}
\usepackage{siunitx}
\usepackage{upgreek}
\usepackage{tabularx} 
\usepackage[utf8]{inputenc}
\usepackage[T1]{fontenc}
\usepackage{mathptmx}
\usepackage{ifthen}
\usepackage{mathtools}
\usepackage[draft]{changes}
\setdeletedmarkup{\color{red}\sout{#1}}

\newcommand{\pd}[2]{\frac{\partial #1}{\partial #2}}

\newcommand{\brackvec}[3]{\begin{bmatrix*}[l]
#1 \\
#2 \\
#3
\end{bmatrix*}}

\definechangesauthor[color=green  , name={Connor E. Murphy}    ]{CM}
\definechangesauthor[color=blue   , name={Tahereh Naderishahab}]{TN}
\definechangesauthor[color=red    , name={Brian D'Urso}        ]{BD}
\definechangesauthor[color=orange , name={Cody Jessup}         ]{CJ}
\definechangesauthor[color=yellow , name={Max Fields}         ]{MF}
\definechangesauthor[color=purple , name={Zach Etienne}        ]{ZE}
\definechangesauthor[color=magenta, name={Leo Werneck}         ]{LW}

\makeatletter
\def\@email#1#2{%
 \endgroup
 \patchcmd{\titleblock@produce}
  {\frontmatter@RRAPformat}
  {\frontmatter@RRAPformat{\produce@RRAP{*#1\href{mailto:#2}{#2}}}\frontmatter@RRAPformat}
  {}{}
}%
\makeatother
\begin{document}

\preprint{}

\title{Ultra-low damping of the translational motion of a composite graphite rod in a magneto-gravitational trap}
\author{Connor~E.~Murphy}
\affiliation{Department of Physics, Montana State University, Bozeman, Montana 59717, USA}

\author{Cody~Jessup}
\affiliation{Department of Physics, Montana State University, Bozeman, Montana 59717, USA}

\author{Tahereh~Naderishahab}
\affiliation{Department of Physics, Montana State University, Bozeman, Montana 59717, USA}

\author{Yateendra~Sihag}
\affiliation{Department of Physics, Montana State University, Bozeman, Montana 59717, USA}

\author{Max~M.~Fields}
\affiliation{Department of Physics, Montana State University, Bozeman, Montana 59717, USA}

\author{Leonardo~R.~Werneck}
\affiliation{Department of Physics, University of Idaho, Moscow, Idaho 83844, USA}

\author{Zachariah~B.~Etienne}
\affiliation{Department of Physics, University of Idaho, Moscow, Idaho 83844, USA}
\affiliation{Department of Physics and Astronomy West Virginia University, Morgantown, West Virginia 26506, USA}
\affiliation{Center for Gravitational Waves and Cosmology,
Chestnut Ridge Research Building,
Morgantown, West Virginia 26506, USA}

\author{Brian~D'Urso\textsuperscript{*}}
\email{durso@montana.edu}
\affiliation{Department of Physics, Montana State University, Bozeman, Montana 59717, USA}

\date{\today}

\begin{abstract}
We demonstrate an ultra-low dissipation, one-dimensional mechanical oscillator formed by levitating a millimeter-scale composite graphite rod in a room-temperature magneto-gravitational trap. The trap's magnetic field geometry, based on a linear quadrupole, eliminates first-order field gradients in the axial direction, yielding a low oscillation frequency with ultra-low eddy-current losses. Direct ring-down measurements under vacuum compare the damping of the vertical and axial motion; while the vertical motion damps in seconds, the axial motion damps with a time constant of over 5 days. Analysis reveals that this dramatic difference in damping is a result of the symmetry of the magnetic field and the anisotropy of the trap strength. The results are remarkably robust, demonstrating a potential platform for inertial and gravitational sensing.
\end{abstract}

\maketitle
\sloppy 

Levitated optomechanical systems consisting of an optically-trapped silica nanosphere have been used to demonstrate cooling to the quantum ground state of a kilohertz harmonic oscillator by taking advantage of the low damping in these systems.\cite{delic2020cooling,tebbenjohanns2021quantum,magrini2021real,xu2024simultaneous} For inertial and gravitational sensing, systems with larger masses and lower oscillation frequencies can be beneficial.\cite{lewandowski2021high,monteiro2020force,li2023collective,WANG2020112122,RademacherMillenQuantumSensing} Levitated optomechanical systems have also been employed for dark matter searches.\cite{monteiroSearchforDarkMatter2020,KilianDarkMatter} To minimize noise, the low damping and isolation from the surrounding environment of levitated systems are critical.\cite{timberlake2019acceleration} In particular, magneto-gravitational traps (MGTs) have been used to levitate diamagnetic particles ranging in size from about \qty{1}{\micro \meter}\cite{hsu2016cooling,slezak2018cooling} to \qty{300}{\micro \meter}.\cite{doi:10.1063/1.5051667}

MGTs offer an especially attractive solution for creating low-frequency, large mass harmonic oscillators for inertial and gravitational sensing due to their static trapping mechanism and ability to trap large masses. An MGT geometry similar to the one discussed here has previously been used to levitate charged \qty{50}{\micro \meter} borosilicate spheres with the aid of a vertical electric field.\cite{lewandowski2021high} However, that system has high damping rates due to the movement of mirror charges in the conductive pole pieces.\cite{brown1986ChargeDamping}

While the diamagnetism of superconductors can be used to levitate a variety of magnetic materials, such as ferrimagnetic yttrium iron garnet,\cite{fuwaFerromagneticLevitation} in a cryogenic environment, graphite is a promising material for direct magnetic levitation due to its relatively strong diamagnetic susceptibility at room temperature. However, graphite is also conductive, giving rise to eddy-current damping when it moves through a magnetic field gradient.\cite{chen2020rigid}

Levitated oscillators of mass ${\sim}${0.35}--\qty{200}{\milli\gram} using a two-by-two alternating cube, cylindrical, or two-layer permanent magnet array and graphite plates have been constructed with frequencies ranging from {0.6}--\qty{35}{\hertz}\cite{chen2022diamagnetic,xie2023suppressing,tian2024feedback,hart2024characterizing,chen2024nonlinear,wang2025diamagnetically,chen2025levitated,long2025diamagnetically} and damping times ranging from about 0.4--\qty{26}{\second} for pyrolytic graphite (${\sim}$0.35--\qty{60}{\milli\gram}).\cite{chen2024nonlinear,hart2024characterizing,wang2025diamagnetically,long2025diamagnetically} Engineering slits in the graphite has been shown to decrease the eddy-current damping of levitated graphite (${\sim}150$--\qty{200}{\milli\gram}) by more than an order of magnitude, allowing damping times of about {23}--\qty{66}{\second}.\cite{xie2023suppressing,romagnoli2023controlling} The use of graphite-composite resonators (silica-coated or fabricated using powder mixed with wax or epoxy with a mass of ${\sim}${0.4}--\qty{40}{\milli\gram}) has demonstrated even longer damping times of {6}--\qty{22}{\minute}.\cite{chen2022diamagnetic,tian2024feedback} Still longer damping times of hours or even days have been reported for rotational motion of graphite rotors (\qty{191}{\milli\gram}) levitated over cylindrically symmetric permanent magnets,\cite{chen2025levitated,kim2025magnetically} creating low-loss rotors rather than oscillators. Finally, on-resonant detection of vibration modes of levitated graphite resonators (\qty{22}{\milli\gram}) has also been measured, exhibiting kilohertz oscillation frequencies and damping times on the order of tens of milliseconds.\cite{chen2021diamagnetically}

\begin{figure}[ht!] 
\centering \includegraphics[width=1.0\columnwidth]{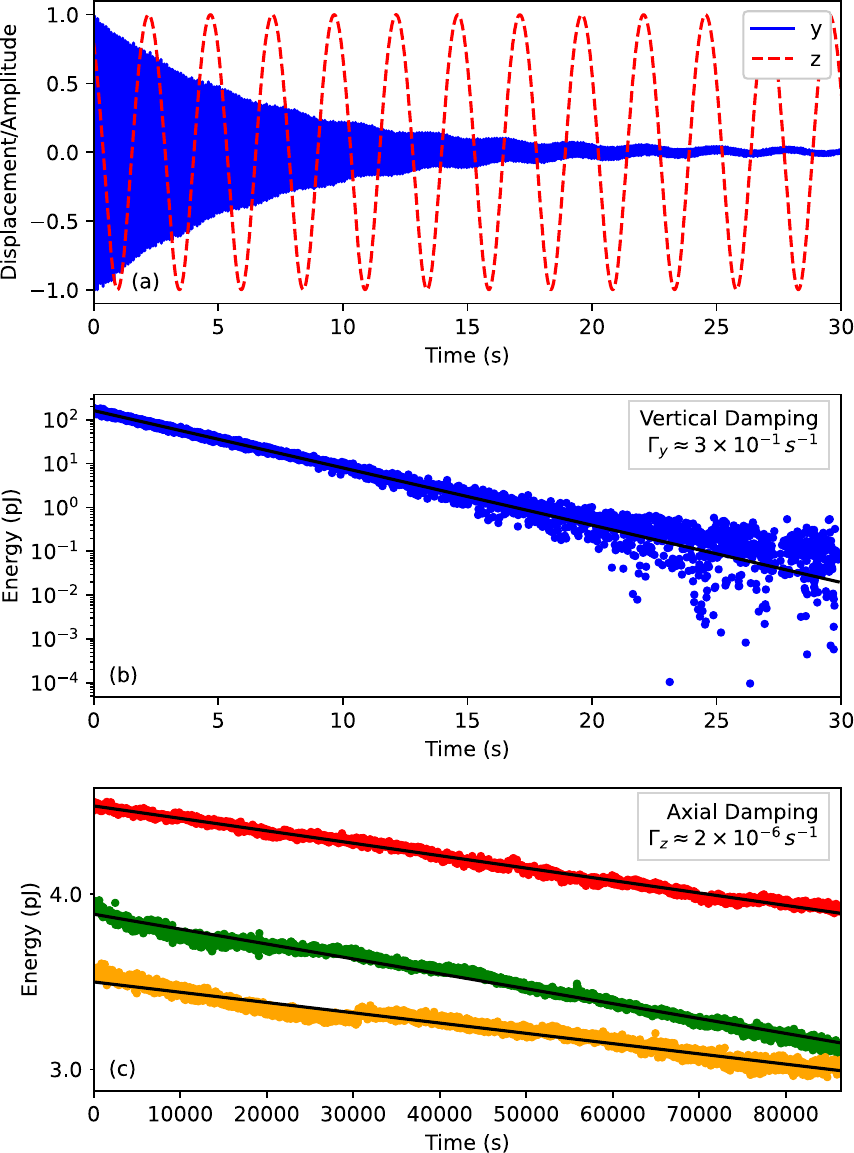}
\caption{\label{fig:ringdowns} Plot (a) displays several oscillation cycles in the vertical ($y$) motion and a few cycles in the axial ($z$) motion of the levitated 4B graphite rod. Plot (b) presents the kinetic-energy fit to the rod’s $y$-motion over a 30-second interval, representative of the five data sets that show consistent behavior, and plot (c) shows the kinetic-energy fit to the rod's $z$-motion over three 24-hour data sets. The exponential fits to the energy are displayed as black lines. The vertical and axial damping rates were measured to be $\Gamma_y = \qty{0.305\pm0.001}{s^{-1}}$ and \mbox{$\Gamma_z = \qty{2.1\pm0.2d-6}{s^{-1}}$}. Only every tenth measured and fitted point is displayed in (c) to reduce the image size.}
\end{figure}

The approach taken here combines the method of using a composite graphite rod with a MGT geometry that effectively creates a one-dimensional, \qty{0.7}{\milli\gram} harmonic oscillator with ultra-low dissipation. In particular, although the vertical motion of the levitated graphite still damps within seconds, the axial mode exhibits an ultra-low damping rate of $\Gamma_z = \qty{2.1\pm0.2d-6}{s^{-1}}$ (Fig.~\ref{fig:ringdowns}) or a damping time of over 5 days. With a low oscillation frequency of \SI{0.40}{\hertz}, this results in a quality factor of about \qty{d6}{}.

\begin{figure}[ht!] 
  \centering
  \includegraphics[width=0.85\columnwidth]{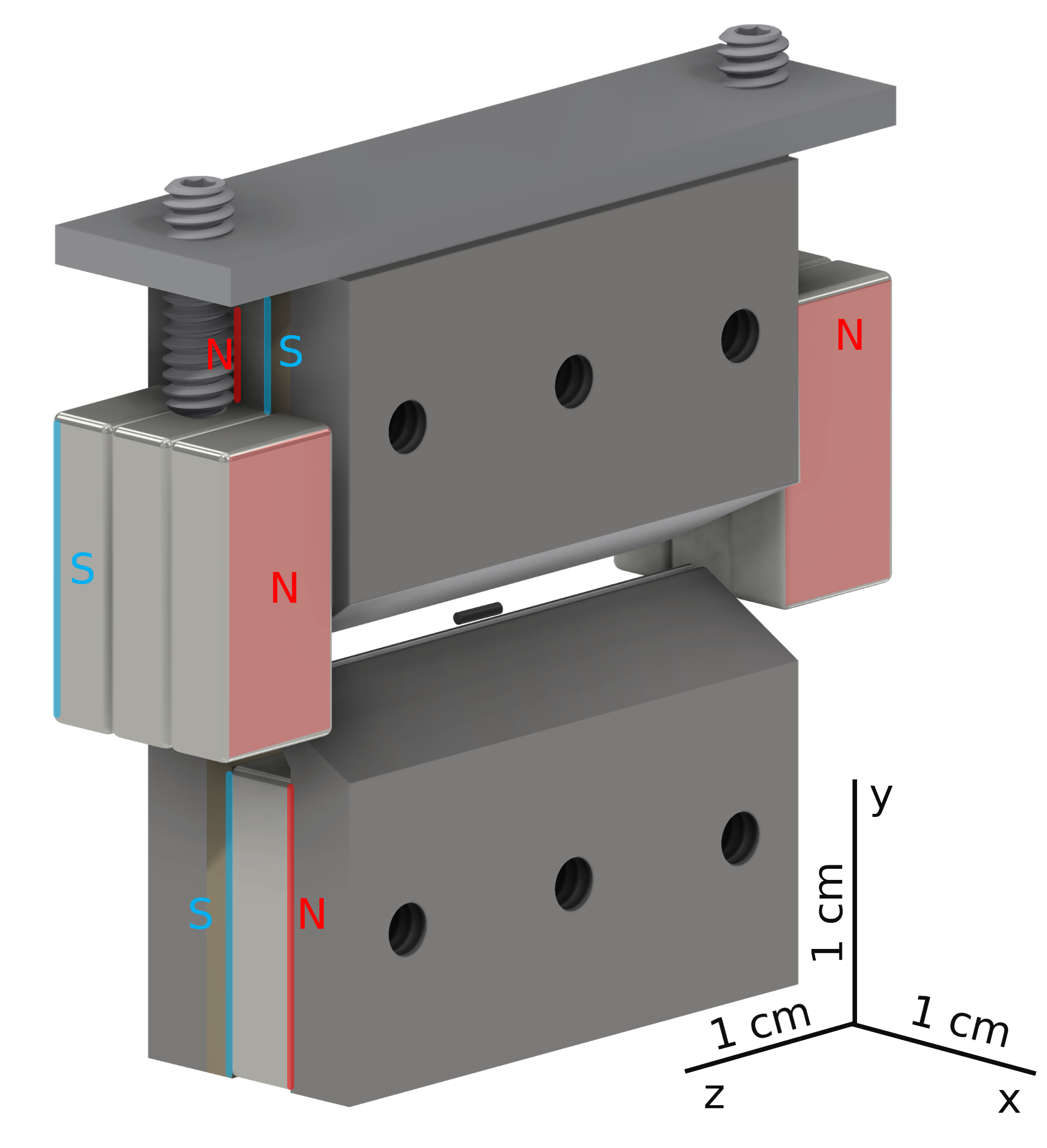}
  \caption{A schematic of the magneto-gravitational trap used for levitating a graphite rod. Set screws are used to adjust the height of the end-magnet stacks (each composed of three smaller NdFeB magnets) on both axial ends of the trap in order to create a low-frequency axial trapping potential. North (N) and south (S) pole labels are shown to describe the magnet orientations relative to each other. The dark cylinder in the gap between the pole pieces represents the trapped composite graphite rod.}
  \label{fig:trap}
\end{figure}

The magneto-gravitational trap used in this experiment is made by sandwiching two nickel-plated NdFeB magnets between four Hiperco~50A (iron-cobalt alloy) pole pieces, as shown in Fig.~\ref{fig:trap}, to create a magnetic field which is dominated by a linear quadrupole component. This results in a strong trapping force in the transverse ($x$) and vertical ($y$) directions, and a weak trapping force in the axial ($z$) direction. Furthermore, there is reason to expect that the damping in the $z$-direction could be weak because the symmetry of the trap results in no linear magnetic field gradients in the $z$-direction at the center of the trap.

Using notation from a multipole expansion of the trap magnetic field\cite{hsu2016cooling} inside an MGT that takes advantage of the symmetries of the system, we write the magnetic field as 
\begin{equation} \label{eq:B}
\vec{B} = -\brackvec{\frac{1}{4} \frac{a_2}{y_0} \sqrt{\frac{15}{\pi}} y + \frac{1}{12} \frac{a_3}{y_0^2} \sqrt{\frac{21}{2\pi}} \left(4z^2 - y^2 - 3x^2\right)}{\frac{1}{4} \frac{a_2}{y_0} \sqrt{\frac{15}{\pi}} x - \frac{1}{6} \frac{a_3}{y_0^2} \sqrt{\frac{21}{2\pi}} xy}{\frac{2}{3} \frac{a_3}{y_0^2} \sqrt{\frac{21}{2\pi}} xz},
\end{equation}
where $a_2$ and $a_3$ parameterize the strength of the quadrupole and hexapole terms, respectively, and $y_0 = \qty{6.14d-4}{\meter}$ is a dimensional scaling factor representative of the trap dimensions. Higher-order terms have been dropped because they are not important for the motion studied here.

\begin{figure}[ht!] 
\centering \includegraphics[width=1.0\columnwidth]{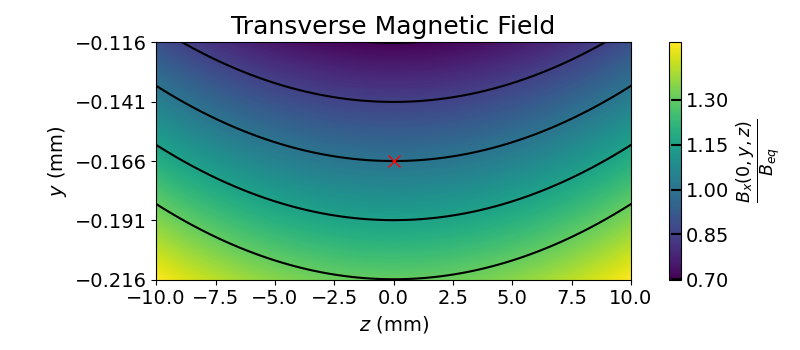}
\caption{\label{fig:Bx} A plot of the relative transverse magnetic field, $B_x(x=0)$, near the rod's equilibrium position (shown by the red ``x''). Note the difference in scale between $z$ and $y$, indicative of the high level of anisotropy in the trap. The $y$-axis is shown magnified 100 times relative to the $z$-axis.}
\end{figure}

A diamagnetic rod in a magnetic field and subject to the Earth's gravitational acceleration $g$ has potential energy
\begin{equation} \label{eq:trapU}
U = - \frac{\chi m}{2 \mu_0 \rho} B^2 - mg y,
\end{equation}
where $\mu_0$ is the permeability of free space, $\chi$ is the volume magnetic susceptibility of the rod, $\rho$ is its mass density, and $m$ is its mass. When trapped, the contribution from gravity gives rise to an offset in the vertical equilibrium position, $y_{\rm eq}$, which in our system is empirically determined to be $y_{\rm eq} = -\qty{0.166}{\milli \meter}$. Trapping in the $z$-direction is a result of the combination of this gravitational offset and (for $x=0$) the $x$-component of the magnetic field, which is plotted in Fig.~\ref{fig:Bx}.

The angular frequency of harmonic oscillation along each trap axis $\zeta = x$, $y$, or $z$ is given by\cite{hsu2016cooling}
\begin{equation} \label{eq:omega}
\omega_\zeta^2 = \frac{-\chi}{2 \rho \mu_0} \frac{\partial^2 B^2}{\partial \zeta^2},
\end{equation}
which for $x=0$ and $y\rightarrow y_{\rm eq}+y$ gives

\begin{align}
\omega_y^2 &\approx \frac{-\chi}{2 \rho \mu_0} \left(\frac{15}{8 \pi} \frac{a_2^2}{y_0^2} -\frac{315}{8\pi} \frac{a_2 a_3}{y_0^3} y_{\rm eq} \right), \label{eq:omega_y}\\
\omega_z^2 &\approx \frac{-\chi}{2 \rho \mu_0} \left(\frac{\sqrt{70}}{2\pi} \frac{a_2 a_3}{y_0^3} y_{\rm eq} \right), \label{eq:omega_z}
\end{align}
where we have neglected terms of order $\zeta^2$ and higher. Since $\omega_z \ll \omega_y$, we drop the second term in Eq.~\eqref{eq:omega_y} in further analysis. In this case, the ratio of $a_3$ to $a_2$ is given by $\frac{a_3}{a_2} = \frac{3\sqrt{70}}{56} \frac{y_0}{y_{\rm eq}}\frac{\omega_z^2}{\omega_y^2} \approx \qty{1.7d-4}{}$, using values of $\omega_y$ and $\omega_z$ measured below. This ratio is used in conjunction with Eq.~\eqref{eq:B} to create the plot shown in Fig.~\ref{fig:Bx}, which displays the relative magnitude of the trap B-field.

By combining Eqs.~\eqref{eq:B}, \eqref{eq:omega_y}, and \eqref{eq:omega_z} and dropping all but the lowest-order terms in each partial derivative, we note that $\pd{B_x}{y}$ is a constant and can write $\pd{B_x}{z}$ as:
\begin{equation} \label{eq:bx-z}
\pd{B_x}{z} = \left(\frac{\omega_z}{\omega_y} \right)^2 \left(\frac{z}{y_{\rm eq}} \right) \pd{B_x}{y}.
\end{equation}
Other components of the field do not contribute a restoring force since $B_y = B_z = 0$ for $x=0$. Critically, for low damping in the $z$-direction, the appearance of the ratio of the frequencies squared is a result of the lack of a linear field gradient in the $z$-direction at the equilibrium position.


To estimate the effect of eddy currents within the rod due to its motion in the $B_x$ field gradients, we model the rod as a loop of wire in the $yz$-plane with resistance $R$ and transverse cross-sectional area $A_x$.

Assuming that the oscillations are weakly damped ($\Gamma_\zeta \ll \omega_\zeta$), the eddy-current damping rate for the system can then be estimated from the ratio of the Ohmic power loss due to induced current $I$ to the energy of the oscillator $E$ averaged over one period as
\begin{equation} \label{eq:gamma}
\Gamma_\zeta = \frac{\langle I^2 R \rangle_\zeta}{\langle E \rangle_\zeta} = \frac{A_x^2}{R} \frac{\left \langle \pd{B_x}{\zeta}^2 \dot{\zeta}^2 \right \rangle_\zeta}{m \omega_\zeta^2 \zeta_{\rm rms}^2}.
\end{equation}
where $\zeta_{\rm rms}$ is the root mean square (rms) of the $x$, $y$, or $z$ displacement from equilibrium.

Eq.~\eqref{eq:gamma}, as discussed in more detail in Supplementary Information Sec.~D, gives the vertical damping rate in the system as
\begin{equation} \label{eq:gamma-y}
\Gamma_y = \left(\pd{B_x}{y}\right)^2 \frac{A_x^2}{m R},
\end{equation}
and  we can determine the axial damping rate by combining Eq.~\eqref{eq:bx-z} with Eq.~\eqref{eq:gamma} to get:
\begin{equation} \label{eq:gamma-z}
\Gamma_z = \left(\pd{B_x}{y}\right)^2 \left(\frac{\omega_z}{\omega_y}\right)^4 \frac{A_x^2 \ z_{\rm rms}^2}{2 m R y_{\rm eq}^2}.
\end{equation}

Finally, the relationship between $\Gamma_z$ and $\Gamma_y$ is
\begin{equation} \label{eq:gamma-ratio}
\Gamma_z = \left(\frac{\omega_z}{\omega_y}\right)^4 \frac{z_{\rm rms}^2}{2 y_{\rm eq}^2} \Gamma_y,
\end{equation}
The appearance of the ratio of the oscillation frequencies to the fourth power suggests the possibility of dramatically lower damping in the weakly-confined axial direction when compared to the much more strongly-confined vertical direction.

\begin{figure}[ht!] 
\centering \includegraphics[width=1.0\columnwidth]{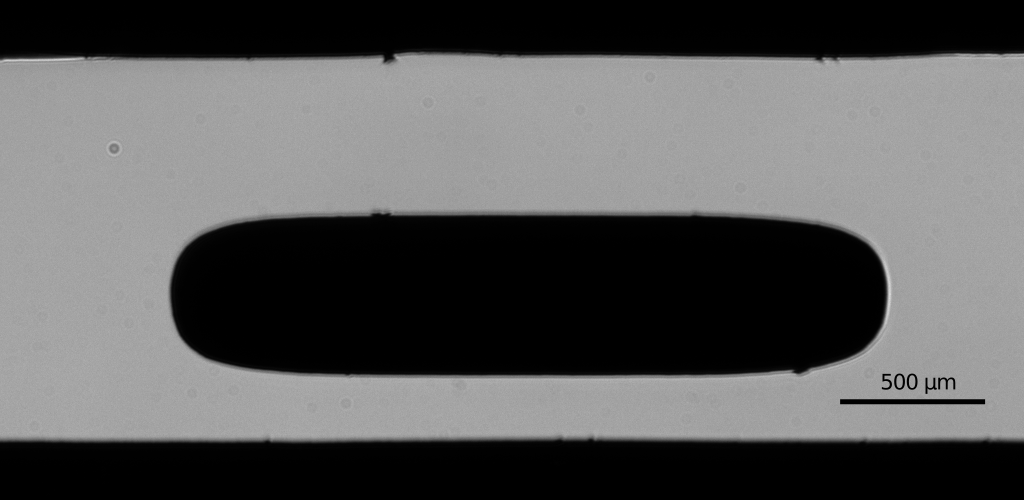}
\caption{\label{fig:4Brod} An image of a piece of ${\sim}\qty{0.5}{\milli \meter}$-diameter piece of Pentel 4B pencil lead levitated in an MGT imaged along the $x$-axis. The top and bottom black sections are the top and bottom pole pieces. The pixel size is calibrated to be \qty{3.45}{\micro\meter} before the images undergo $4 \text{x} 4$ binning when saved.}
\end{figure}

The levitated test mass is a \qty{2.5}{\milli\meter}-long segment of \qty{0.5}{\milli \meter}-diameter Pentel 4B pencil lead, shown in Fig.~\ref{fig:4Brod}. The composition of the pencil lead is approximately \qty{79}{\percent} graphite, \qty{15}{\percent} clay, and \qty{5}{\percent} wax,\cite{sousa2000observational} which likely increases its resistivity compared to pure graphite.\cite{Jain2023resistivity} Prior to levitation, the rod was tumbled,\cite{belsonTumbler} producing rounded ends that avoid the focal-plane ambiguity of flat ends.

Multiple precautions are taken to minimize electrostatic sources of damping. First, the tips of the trap's pole pieces are covered with \qty{25}{\micro \meter}-thick copper foil to decrease surface patch potentials. Second, an Am-241 ionizing radiation source is directed at the trapping region for several minutes at atmospheric pressure to neutralize any net charge accumulated on the rod and adjacent surfaces. The axial trapping potential is then adjusted by moving two end-magnet stacks at the axial ends of the trap (see Fig.~\ref{fig:trap}), which allows for fine-tuning of the axial frequency and ensures a stable potential well.

The position of the rod in the $yz$-plane is recorded by capturing images of the rod back-illuminated with light from a pulsed blue LED and imaged onto a CMOS camera (Basler acA2440-75um) with a Mitutoyo 1X telecentric objective lens. Cross-correlation based image analysis\cite{lewandowski2021high,werneck2024crosscorrelation} is used to precisely determine the apparent rod displacement. For axial (vertical) motion analysis, the recorded frame rate is \qty{500}{\hertz} (\qty{550}{\hertz}) with an illumination time of \qty{25}{\micro\second} in each frame.
 

To decrease damping of the rod's motion due to residual gas, the vacuum chamber enclosing the trap is baked at ${\sim}\qty{100}{\celsius}$ under high vacuum, followed by pumping with ion and titanium sublimation pumps. The pressure in the chamber varied on average from approximately \qtyrange{3d-9}{9d-9}{Torr} over the course of the experiments, as recorded on the ion pump controller.

The vacuum chamber and optics are built on a platform that is suspended on four air isolation mounts for vibration isolation. The axial tilt of the platform is actively stabilized to an rms variation of \qty{1.1}{\micro \radian} to prevent the rod from drifting.\cite{lewandowski2020active}


We analyze multiple recordings of the rod's motion. In the axial direction, three 24-hour data sets of motion are recorded after the rod is excited to an amplitude of approximately {1.2}--\qty{1.4}{\milli\meter} by manually tilting the table on resonance. In the vertical direction, five data sets are recorded where the motion of the rod is excited by striking the vacuum chamber with a rubber mallet, resulting in an initial amplitude of approximately {20}--\qty{30}{\micro\meter}, but the data is only recorded for \qty{60}{\second} due to the rapid damping of the vertical motion. The short damping time of the vertical motion does not permit time for the axial tilt to restabilize; recording after axial excitation did not begin until after the axial tilt had restabilized.

The damping rate for each axis of motion was calculated by finding the kinetic energy of the motion in each direction as a function of time after excitation. This was done by fitting the data points surrounding each zero-crossing in displacement to a cubic function, then analytically differentiating to obtain a velocity (see Supplementary Information Sec.~F). Fig.~\ref{fig:ringdowns} illustrates the motion of the rod as well as exponential fits to the kinetic energy in each degree of freedom over time.
Note the drastic difference in damping rates between vertical and axial motions --- the former damps in a matter of seconds ($\Gamma_y = \qty{0.305\pm0.001}{s^{-1}}, Q_y = \omega_y/\Gamma_y = \num{803\pm3}$) while the latter takes days to lose significant energy ($\Gamma_z = \qty{2.1\pm0.2d-6}{s^{-1}}, Q_z = \omega_z/\Gamma_z = \qty{1.2\pm0.1d6}{}$).


We can use the system parameters and Eq.~\eqref{eq:gamma-ratio} to estimate the axial damping rate due to eddy currents in the rod from the measured vertical damping rate. We measured $z_{\rm rms}\approx\qty{9d-4}{\meter}$ from the axial damping data and $y_{\rm eq} = \qty{-1.7d-4}{\meter}$ from Fig.~\ref{fig:4Brod}. By interpolating around the peak values in the Fourier domain (see Supplementary Information Sec.~C), the vertical and axial frequencies of motion were determined to be $\omega_y/(2\pi)=\qty{39.0}{\hertz}$ and $\omega_z/(2\pi)=\qty{0.40}{\hertz}$. Combining these values together gives $\Gamma_z \approx \qty{5d-8}{\second^{-1}}$, corresponding to a $1/\mathrm{e}$ time over 200 days. 

Since the observed axial damping rate is higher than the estimated value from eddy currents in the rod alone, other sources of damping may be dominant. First, residual charge on the particle would cause damping as a result of the movement of mirror charges in the conductive pole pieces;\cite{brown1986ChargeDamping} the charge on the particle here was not measured and is difficult to estimate. Second, there are eddy currents in the pole pieces due to the perturbation of the magnetic field by the moving rod,\cite{matsko2020on} but these are expected to be small because the magnitude of the magnetic susceptibility of the rod is much less than 1. Third, some damping by residual gas in the chamber is inevitable. The theoretical gas damping rate for a sphero-cylindrical particle like the one shown in Fig.~\ref{fig:4Brod} at the pressures measured is in the range \qtyrange{3d-9}{9d-9}{\second^{-1}} (see Supplementary Information Sec.~E) and scales linearly with pressure.\cite{cavalleri2010gas} Lastly, there could be some energy exchange between the other degrees of freedom as suggested by the presence of apparent vertical motion at the axial frequency in Fig.~\ref{fig:ringdowns}(a) and Fig. S5 in Supplementary Information. This could be due to a tilt in the camera or irregularities in the pole piece surfaces causing bumps in the potential. 

Eq.~\eqref{eq:gamma-ratio} shows how the field geometry of the MGT makes it possible to obtain ultra-low damping in one trap axis coincident with strong damping in the other two orthogonal axes (the transverse motion is not analyzed but is similar to the vertical motion). Similarly, rotational or librational degrees of freedom are expected to be highly damped. Having ultra-low dissipation in the axial direction combined with strong dissipation in the other degrees of freedom damps any non-axial motion of the rod rapidly, effectively creating a one-dimensional oscillator without the need for active feedback cooling of the other degrees of freedom.\cite{tian2024feedback}

It is interesting to note that these results were obtained without careful fine-tuning of the alignment of the pole pieces --- similar results were obtained with three different instances of assembly of the trap by two different people as well as multiple Pentel graphite pencil lead rods of two different compositions. This indicates a low-damping result that is remarkably robust to misalignment, although it is possible that eddy currents in the rod still dominate the axial damping. Data from another particle with a different composition, shape, trap build, pressure, and oscillation frequency are shown in the Supplementary Information; in that case, external sources of excitation appeared to overwhelm the remaining damping, but the ultimate limits that can be placed on the damping are similar to those from the data in Fig.~\ref{fig:ringdowns}.

Further improvements on the design might be made by adjusting the locations of the end-magnet stacks to further flatten the potential, by implementing some other form of potential tuning,\cite{long2025diamagnetically} and by improving the isolation of the system from external disturbances.

\section*{Supplementary Information}

Additional details are provided on similar results from an HB Pentel graphite rod; eddy-current damping calculations; estimates of other damping or excitation channels (residual gas and external vibrations); trap construction and dimensions; and the techniques used to extract peak velocities from position data for kinetic-energy analysis.

\begin{acknowledgments}
This work was primarily supported by the National Science Foundation under Grant Nos.~1912083, 2011783, and 2513013 (B.D.), as well as Grant Nos.~2011817, 2227079, and 2513014 (Z.B.E.).
\end{acknowledgments}

\section*{Data Availability Statement}

The data that support the findings of this study are available from the corresponding author upon reasonable request.

\bibliography{bibliography_capfixed}

\end{document}


\preprint{}

\title{Ultra-low damping of the translational motion of a composite graphite rod in a magneto-gravitational trap:\\ Supplementary Information}
\author{Connor~E.~Murphy}
\affiliation{Department of Physics, Montana State University, Bozeman, MT}

\author{Cody~Jessup}
\affiliation{Department of Physics, Montana State University, Bozeman, MT}

\author{Tahereh~Naderishahab}
\affiliation{Department of Physics, Montana State University, Bozeman, MT}

\author{Yateendra~Sihag}
\affiliation{Department of Physics, Montana State University, Bozeman, MT}

\author{Max~M.~Fields}
\affiliation{Department of Physics, Montana State University, Bozeman, MT}

\author{Leonardo~R.~Werneck}
\affiliation{Department of Physics,
University of Idaho,
Moscow, ID}

\author{Zachariah~B.~Etienne}
\affiliation{Department of Physics,
University of Idaho,
Moscow, ID}
\affiliation{Department of Physics and Astronomy,
West Virginia University,
Morgantown, WV}
\affiliation{Center for Gravitational Waves and Cosmology,
Chestnut Ridge Research Building,
Morgantown, WV}

\author{Brian~D'Urso\textsuperscript{*}}
\email{durso@montana.edu}
\affiliation{Department of Physics, Montana State University, Bozeman, MT}


\maketitle

\section*{Supplementary Material}

\subsection{HB Rod Damping}

\begin{figure}[h!]
  \centering
  \includegraphics[width=\columnwidth]{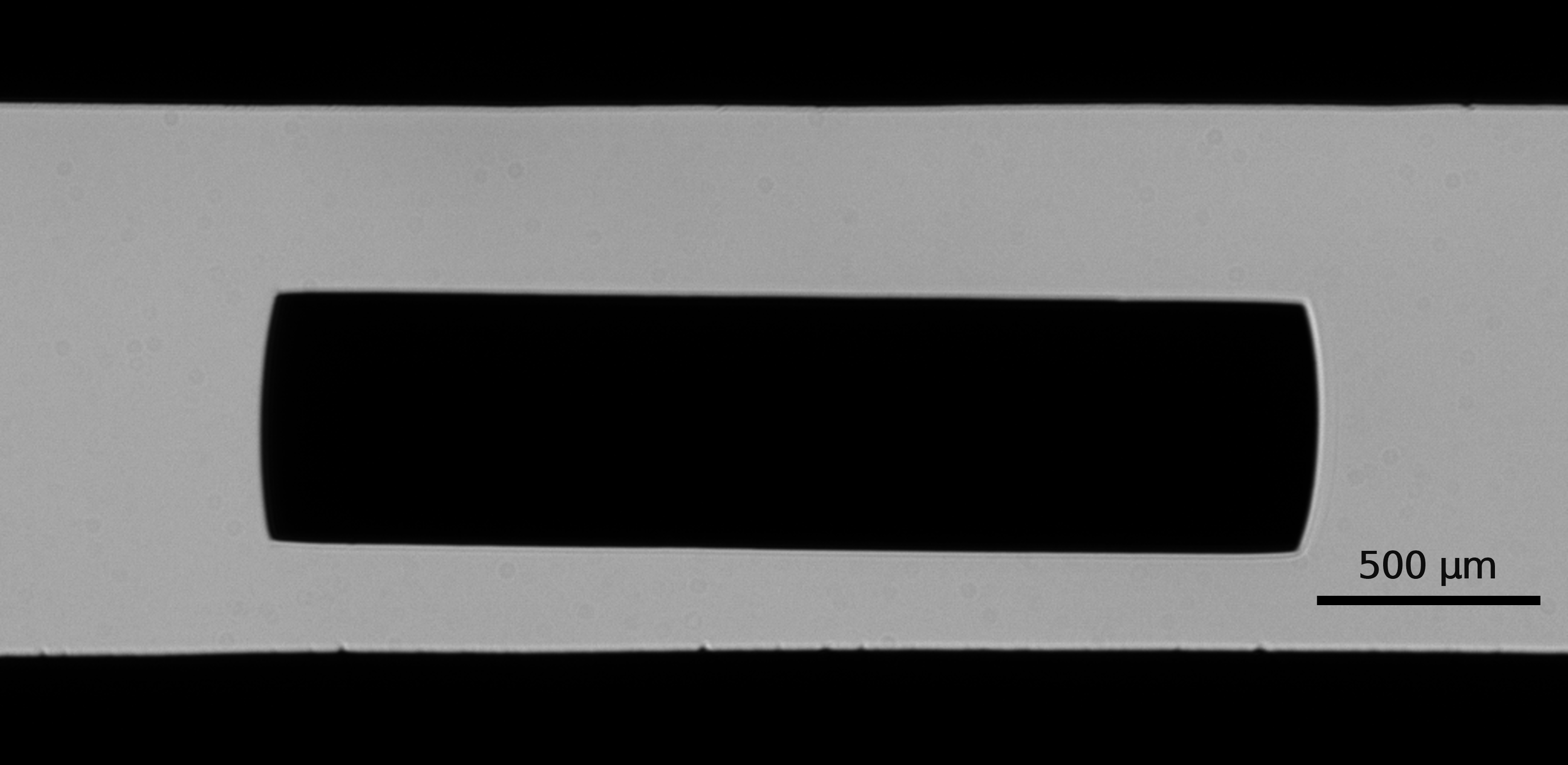}
  \caption{An image along the $x$-axis of a ${\sim}\qty{0.5}{\milli \meter}$-diameter piece of Pentel HB pencil lead (rod) levitated in an MGT. The top and bottom black sections are the top and bottom pole pieces. The pixel size is calibrated to be \qty{3.45}{\micro\meter}.}
  \label{fig:HBparticle}
\end{figure}

The damping of a second graphite composite rod is also measured. The rod consists of a \qty{2.4}{\milli\meter}-long segment of \qty{0.5}{\milli \meter}-diameter Pentel HB pencil lead, as shown in Fig.~\ref{fig:HBparticle}. The composition of HB pencil lead is approximately \qty{68}{\percent} graphite, \qty{26}{\percent} clay, and \qty{5}{\percent} wax.\cite{sousa2000observational} Unlike the 4B pencil lead, the HB pencil lead has ends that are lapped smooth instead of rounded.

The magnetic trap was disassembled and reassembled between experimental runs measuring the HB and 4B rod behavior in the trap. Since the pole piece alignment and positioning of the magnets may not be identical, the oscillation frequencies and damping rates of the HB and 4B rods cannot be directly compared.

\begin{figure}[ht!]
  \centering
  \includegraphics[width=\columnwidth]{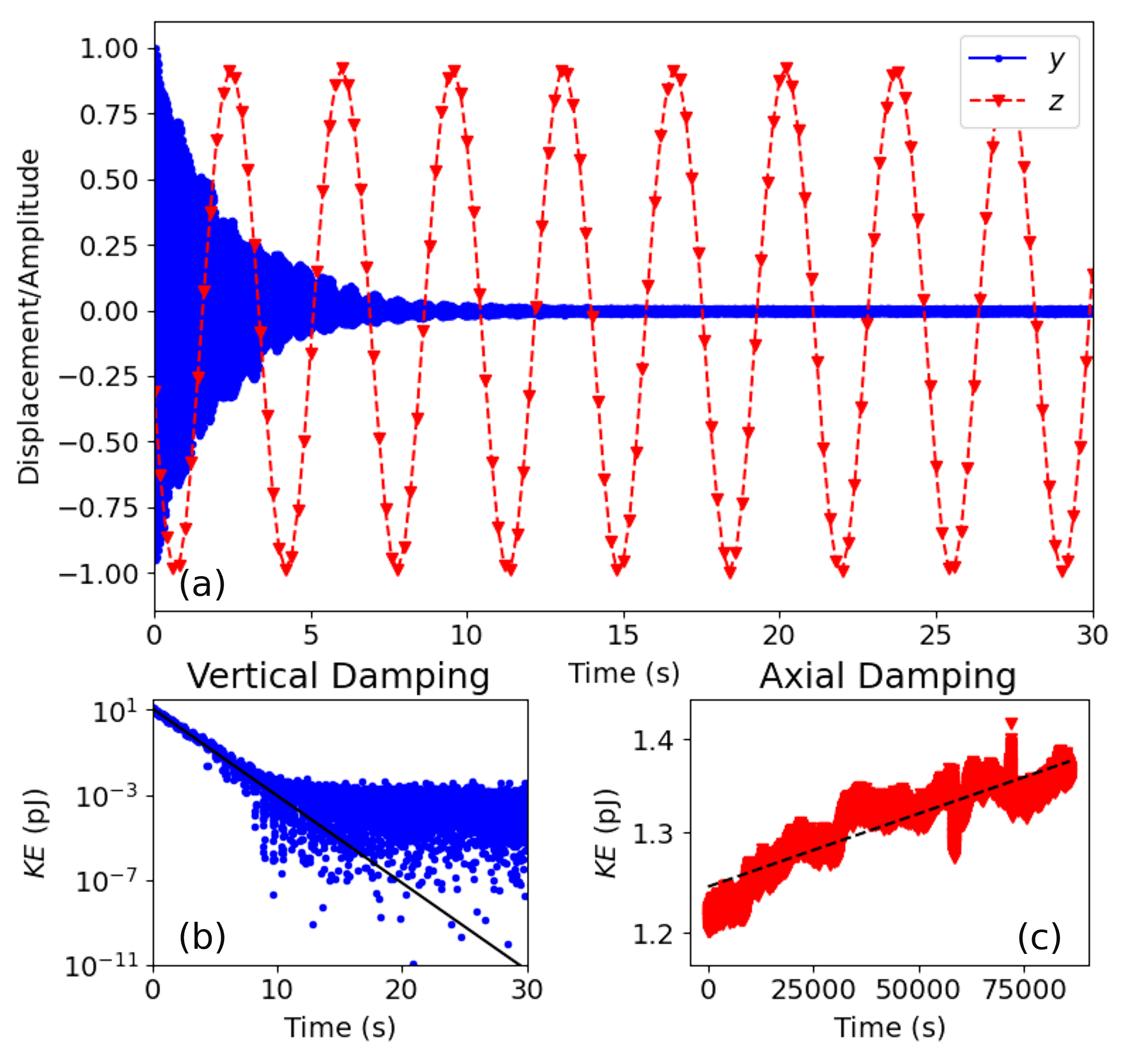}
  \caption{Plot (a) displays several oscillation cycles in the vertical ($y$) motion and a few cycles in the axial ($z$) motion of the levitated HB graphite rod shown in Fig.~\ref{fig:HBparticle}. Plot (b) shows the kinetic-energy fit to the rod's $y$-motion over 30 seconds, and plot (c) shows the kinetic-energy fit to the rod's $z$-motion over 24 hours. The exponential fits to the energy are displayed as black lines. The vertical damping rate was measured to be $\Gamma_y = \qty{0.91\pm0.02}{s^{-1}}$ and the effective axial damping rate, which varied greatly from run to run, was found in this particular measurement to be about $\Gamma_z \approx -\qty{1.2d-6}{s^{-1}}.$ The negative value reflects the slow increase in energy, presumably due to environmental noise sources driving the nearly undamped motion.}
  \label{fig:HB_ringdowns}
\end{figure}

The vertical and axial ringdown data is shown in Fig.~\ref{fig:HB_ringdowns}. The measured damping of the vertical motion of the HB rod was somewhat higher than the 4B rod, while no damping of the axial motion of the HB rod could be detected. Instead, the HB rod slowly gained energy, presumably due to external vibrations or another excitation mechanism. The lack of observable damping in the HB rod could be due to higher vibrations or lower damping than with the 4B rod despite higher gas pressure (see Sec.~\ref{sec:SI_pressure}). Typical external vibrations, recorded during one of the 4B rod trials, is described in Sec.~\ref{sec:SI_vibration}. There is no indication of particularly strong vibration around the axial frequency that might drive the motion.

\subsection{External Vibrations}
\label{sec:SI_vibration}
\begin{figure}[ht!] 
  \centering
  \includegraphics[width=\columnwidth]{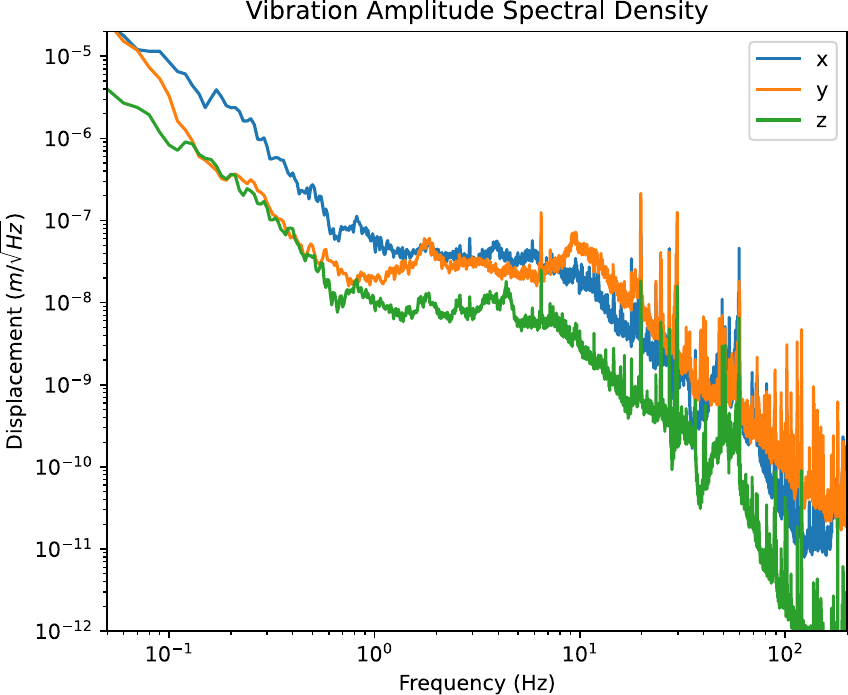}
  \caption{The amplitude spectral density of floor vibrations measured using the SercelInc. L-4C 3D geophone, with the transfer function of the geophone divided out. Due to the approximately \qty{1}{\hertz} resonance of the geophone itself, the ASD may have excess noise below \qty{1}{\hertz}. The ASD is calculated using Welch segmenting with \qty{50}{\percent} overlap with \qty{100}{\second} segments and a Hann window.}
  \label{fig:ASDvibration}
\end{figure}

The amplitude spectral density (ASD) of the vibration of the floor next to the experimental apparatus is shown in Fig.~\ref{fig:ASDvibration}. 
It was recorded using a Sercel Inc.~L-4C 3D geophone, which provides three-dimensional seismic data by sensing ground velocity. The geophone outputs were filtered by Stanford Research Systems (SRS) SRS640 low-pass filters, configured with a cutoff frequency of \qty{200}{\hertz}. This cutoff lies below the Nyquist frequency of \qty{250}{\hertz}. The plot in Fig.~\ref{fig:ASDvibration} represents vibration data in the axial, transverse, and vertical directions from one of the three 24-hour 4B rod axial ringdown data sets. Since the results were very similar across multiple data sets, only one is shown here.

To examine the effects of floor vibrations and determine their magnitude in driving the axial motion of the rod, separate from manually exciting it on resonance, we multiply the ASD of the floor vibrations, $\left|A_z(\omega)\right|$, by the frequency response of the rod's motion with respect to the camera\cite{lewandowski2021high}, yielding
\begin{equation} \label{eq:AxialTransferFunc}
\left|z'(\omega)\right|
= \left|A_z(\omega)\right|\frac{\omega^{2}}
{\left[\,(\omega_z^{2}-\omega^{2})^{2} + \Gamma_z^{2}\omega^{2}\,\right]^{1/2}},
\end{equation}
where $\left|z'(\omega)\right|$ is the ASD of the motion of the rod with respect to the camera. This is plotted as the green line in Fig.~\ref{fig:spectrum_z}. At the axial frequency the amplitude of the rod from floor vibrations is about three orders of magnitude less than the amplitude from manual excitation. Isolation mounts on the experiment likely result in even less vibration at the trap. This analysis was not performed on the vertical motion since additional transient vibrations may have resulted from the process of exciting the vertical motion with a mallet strike.
  
\subsection{Frequency Extraction}

\begin{figure}[ht!] 
  \centering
  \includegraphics[width=\columnwidth]{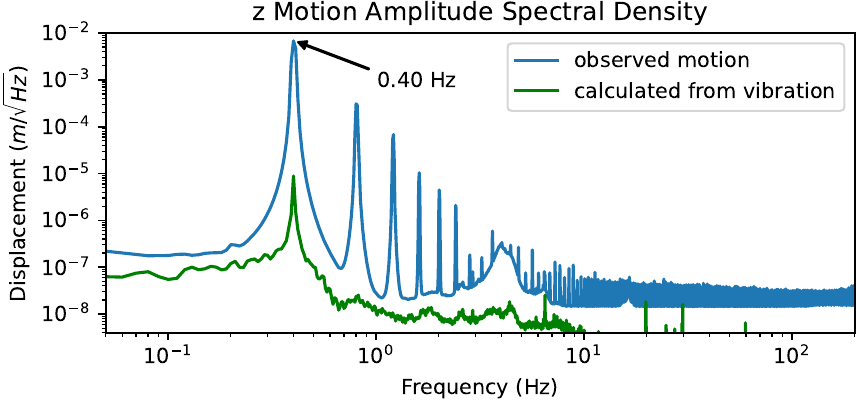}
  \caption{The amplitude spectral density of the 4B rod from one of three 24-hour data sets used to record its $z$-ringdown after being excited to an initial amplitude and the amplitude spectral density of the rod's motion if only driven by measured floor vibration. Both ASDs are calculated using Welch segmenting with \qty{50}{\percent} overlap with \qty{100}{\second} segments and a Hann window. The other data sets gave similar spectra. The measured axial frequency is $\omega_z/(2\pi) = \qty{0.40}{\hertz}$. Note that most of the higher-frequency peaks are harmonics of the fundamental axial frequency due to the anharmonicity of the motion.\cite{Giliberti2014Anharmonicity} The origin of the broader peak at approximately \qty{4}{\hertz} is unknown, but may be due to highly-damped vibrational motion.}
  \label{fig:spectrum_z}
\end{figure}

\begin{figure}[ht!] 
  \centering
  \includegraphics[width=\columnwidth]{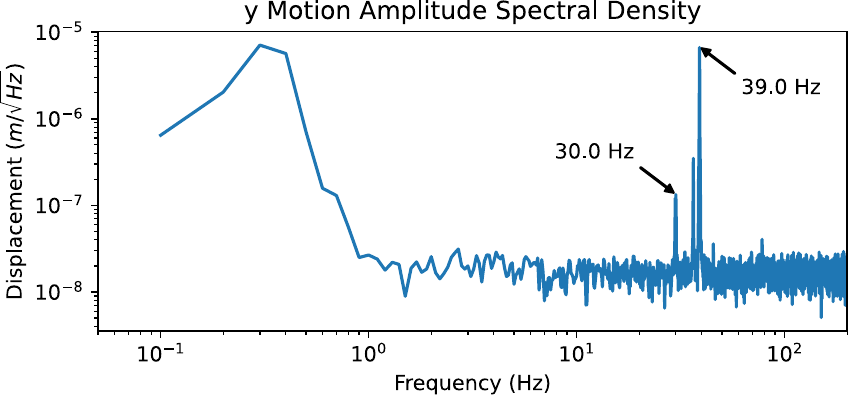}
  \caption{Amplitude spectral density of the 4B rod from one of five \qty{60}{\second} data sets used to record its $y$-ringdown. The ASD is calculated using Welch segmenting with \qty{50}{\percent} overlap with \qty{10}{\second} segments and a Hann window. The other data set gave similar spectra. The measured vertical frequency is $\omega_y/(2\pi) = \qty{39.0}{\hertz}$. The peak at \qty{30.0}{\hertz} is likely due to weak transverse oscillation. The axial resonance peak near \qty{0.4}{\hertz} is also visible. The origin of the weaker peak just below \qty{39}{\hertz} is unknown.}
  \label{fig:spectrum_y}
\end{figure}

The natural frequencies of the oscillator along its three principal axes can be measured from the amplitude spectral density (ASD) of the rod's motion.  The ASD of the $z$ and $y$ motions for the 4B rod are shown in Fig.~\ref{fig:spectrum_z} and Fig.~\ref{fig:spectrum_y}, respectively.

\subsection{Eddy Current Damping}
To get from Eq.~(7) to Eqs.~(8) and (9) in the main text, we can express the long-term behavior of the system as
\begin{align}
\ddot{y} &= - \omega_y^2 y - \Gamma_y \dot{y} \\
\ddot{z} &= - \omega_z^2 z - \Gamma_z(z) \dot{z},
\end{align}
noting that in general $\Gamma_z$ cannot be assumed to be independent of amplitude.

This relationship allows us to estimate the average energy and $I^2R$-losses for $\Gamma_\zeta \ll \omega_\zeta$. The average energy is given by 
\begin{align} \label{eq:E_avg}
\begin{split}
\langle E \rangle_\zeta &= \left \langle \frac{1}{2} m \omega_\zeta^2 \zeta^2 + \frac{1}{2} m \dot{\zeta}^2 \right \rangle_\zeta \\
&\approx \frac{1}{2} m \omega_\zeta^2\left \langle \zeta_0^2 \cos^2 \omega_\zeta t + \zeta_0^2 \sin^2 \omega_\zeta t \right \rangle_\zeta \\
&= m \omega_\zeta^2 \zeta_{\rm rms}^2,
\end{split}
\end{align}
where $\zeta_0$ is the amplitude of the motion at $t\approx0$.

The current induced by the motion of the rod through a field gradient can be calculated using Faraday's law:
\begin{equation} \label{eq:eddy-current}
I = \frac{V}{R} = -\frac{A_x}{R} \fd{B_x}{t} = -\frac{A_x}{R} \pd{B_x}{\zeta} \dot{\zeta}.
\end{equation}

The value $\langle \pd{B_x}{y}^2 \dot{y}^2 \rangle_y$ can trivially be calculated due to the fact that $\pd{B_x}{y}$ is approximately constant. For $z$, however, it is more complicated:
\begin{align} \label{eq:z2zdot2_avg}
\begin{split}
\left \langle \pd{B_x}{z}^2 \dot{z}^2 \right \rangle_z &= \left(\frac{\omega_z}{\omega_y}\right)^4 \frac{1}{y_{\rm eq}^2} \langle z^2 \dot{z}^2 \rangle_z \\
&\approx \left(\frac{\omega_z}{\omega_y}\right)^4 \frac{1}{y_{\rm eq}^2} \left\langle z_0^4 \omega_z^2 \sin^2\omega_zt \ \cos^2\omega_zt \right\rangle_z \\
&= \left(\frac{\omega_z}{\omega_y}\right)^4 \frac{\omega_z^2 z_0^4}{8y_{\rm eq}^2},
\end{split}
\end{align}
which yields the form of Eq.~(9) from the main text using the substitution $z_{\rm rms}^2 \approx \frac{1}{2} z_0^2$.

The value of $y_{\rm eq}$ for the 4B rod is measured to be about \qty{-0.166}{\milli\meter}, while for the HB rod it is about \qty{-0.101}{\milli\meter}. The values of $z_{\rm rms}$ corresponding to the data shown in the top, middle, and bottom curves in Fig.~4(c) in the main text were calculated to be \qty{0.98}{\milli\meter}, \qty{0.95}{\milli\meter}, \qty{0.86}{\milli\meter}, respectively. The value of $z_{\rm rms}$ for the plot shown in Fig.~\ref{fig:HB_ringdowns} is \qty{0.79}{\milli\meter}.

Also of interest for eddy current damping estimations is the resistivity of the graphite composite, which is related to the value of $R$ that appears in Eq.~\eqref{eq:eddy-current}. While a measurement of the resistivity of pencil lead from wood pencils has already been made,\cite{Jain2023resistivity} no measurement of Pentel pencil lead was found in the literature.

To address this, four-wire resistance measurements were performed on 0.5 mm-diameter Pentel HB and 4B pencil leads. For each grade, measurements were taken on 12 different lead segments, and the results were averaged to yield resistivities of \mbox{$1/\sigma_{\mathrm{HB}} = \qty{2.16 \pm 0.07e-6}{\ohm\meter}$} and \mbox{$1/\sigma_{\mathrm{4B}} = \qty{5.38 \pm 0.10e-6}{\ohm\meter}$}. Since the resistivity of the 4B lead is higher than that of HB, this difference cannot explain the observed axial damping rates but may contribute to the difference in vertical damping rates.

\subsection{Theoretical Gas Damping}
\label{sec:SI_pressure}
The gas damping rate for a sphero-cylindrical rod with radius $R=\qty{0.276}{\milli\meter}$ and cylindrical length $h=\qty{1.92}{\milli\meter}$ (volume given by $\frac{4}{3} \pi R^3 + \pi R^2 h$) can be estimated from\cite{cavalleri2010gas} 
\begin{equation}
\Gamma_{\text{sphero-cylinder}} = \frac{p}{\rho_{\rm 4B} \left(R+\frac{3}{4} h\right)} \sqrt{\frac{8 m_0}{\pi k_B T}}\left(1+\frac{\pi}{8} +\frac{3h}{8R}\right),
\end{equation}
where $p$ is the pressure, $m_0 = \qty{4.65d-26}{\kilo\gram}$ was used for the mass of a $N_2$ molecule, $\rho_{\rm 4B} = \qty{1.67\pm0.02d3}{\kilo\gram/\meter^3}$ was directly measured for 4B pencil lead, and $T=\qty{297}{\kelvin}$ was measured to be the approximate ambient temperature. For a pressure range of \qtyrange{3d-9}{9d-9}{Torr}, this corresponds to a damping rate of $\qtyrange{3d-9}{9d-9}{\second^{-1}}$.

For the HB rod used, the damping rate can be found by approximating the rod as a cylinder with radius $R=\qty{0.285}{\milli\meter}$ and length $h=\qty{2.35}{\milli\meter}$, so that the theoretical gas damping rate is given by\cite{cavalleri2010gas}
\begin{equation}
\Gamma_{\rm cylinder} = \frac{p}{\rho_{\rm HB} h} \sqrt{\frac{8 m_0}{\pi k_B T}}\left(1 +\frac{h}{2R}+\frac{\pi}{4}\right),
\end{equation}
where $\rho_{\rm HB} = \qty{1.70\pm0.02d3}{\kilo\gram/\meter^3}$ was directly measured for HB pencil lead. For a pressure range of \qtyrange{3d-8}{6d-8}{Torr}, this corresponds to a damping rate of $\qtyrange{3d-8}{6d-8}{\second^{-1}}$.

Note that the masses of the HB and 4B rods can be estimated from these parameters to be about \qty{1.0}{\milli\gram} and \qty{0.91}{\milli\gram}, respectively (uncertainty is dominated by estimations of volume by approximating rod shape and therefore difficult to estimate). The directly-measured mass of each rod, which was used in calculations of the kinetic energy of the rod, was found to be \qty{0.7\pm0.1}{\milli\gram}.

\subsection{Kinetic Energy Analysis}

To analyze the damping behavior of an oscillator, it is useful to be able to estimate its energy as a function of time. Since the system under study in this experiment is somewhat anharmonic, making a precise calculation of potential energy difficult, it is beneficial to analyze the kinetic energy of the rod over time.

Since the mass of the rod can be considered to be invariant in time, measuring velocity squared is sufficient to characterize the energy damping of the rod's motion.

\begin{figure}[ht!] 
  \centering
  \includegraphics[width=\columnwidth]{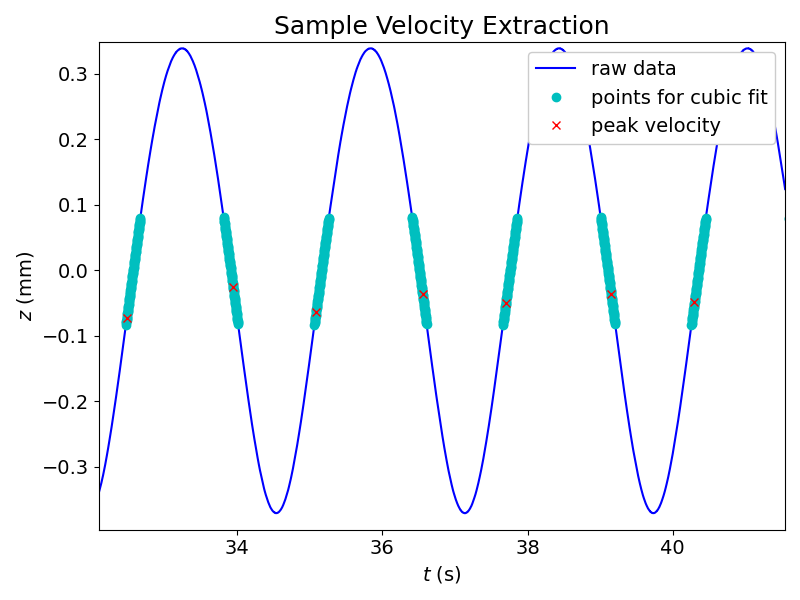}
  \caption{A portion of axial displacement data for the 4B rod shown to demonstrate the method used to calculate the kinetic energy for estimating damping rates. The fits shown use 100 points around each zero-crossing to fit a cubic to the data, allowing analytical differentiation to be used to calculate peak velocities.}
  \label{fig:velocities}
\end{figure}

The peak velocity every half-cycle was determined from the time-domain $z$-motion of the rod by fitting a cubic to a set of points around each zero-crossing, as shown in Fig.~\ref{fig:velocities}. The cubic fit can then be analytically differentiated to obtain a parabolic fit to the velocities, so that the vertex of each parabola estimates the peak velocity of the rod.

The square of these peak velocities can then be plotted in time to estimate the damping rate for the $z$-motion of the rod, as shown in Fig.~4 in the main text.

The same method was used to find the peak velocities for the vertical motion, except that fewer points were used in each fit due to the higher oscillation frequency.

\subsection{Trap Design}

\begin{figure}[ht!] 
  \centering
  \includegraphics[width=\columnwidth]{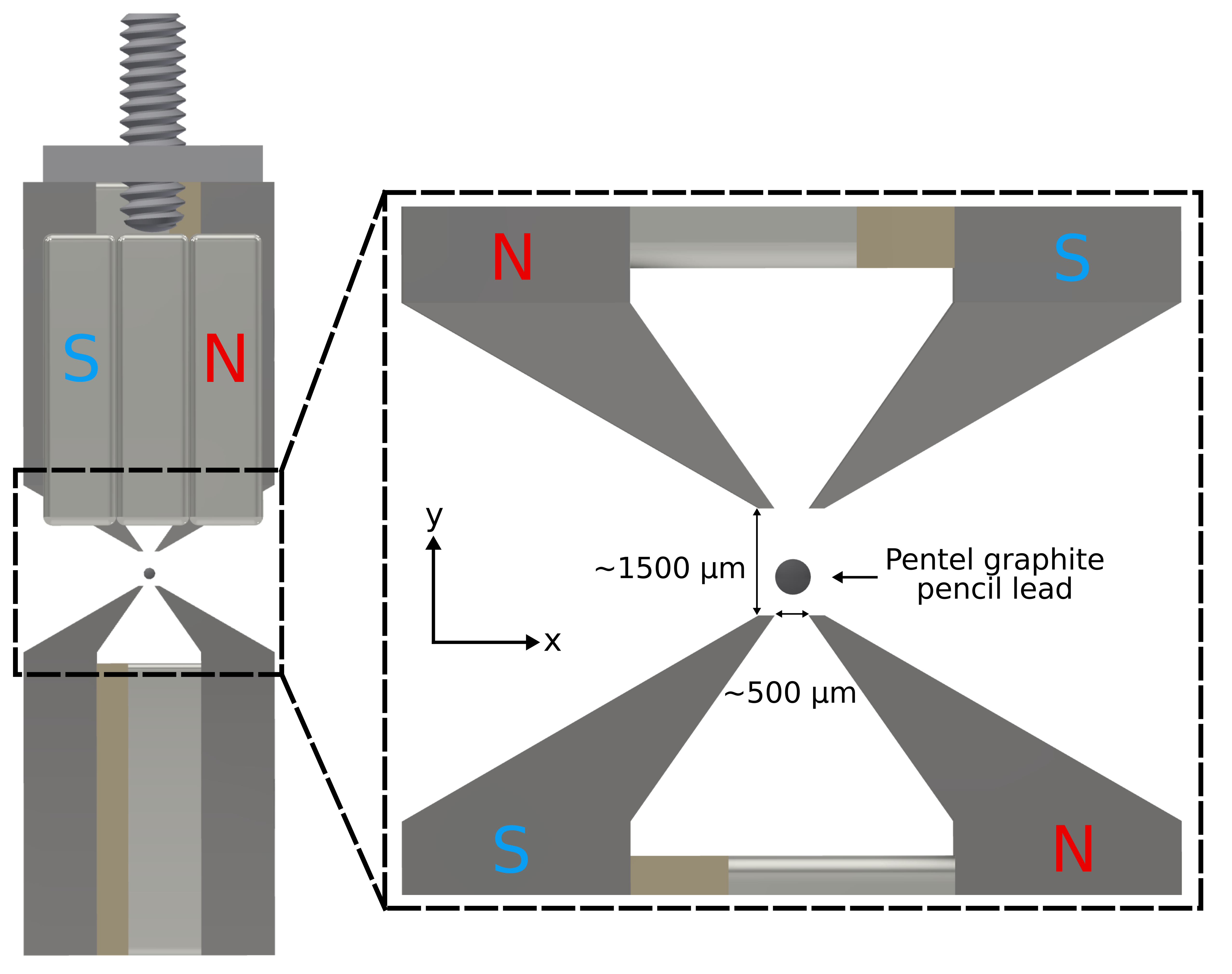}
  \caption{A 3-D scaled drawing displaying the dimensions of the gap sizes between the tips of the pole pieces with respect to the levitating Pentel pencil lead rod (diameter ${\sim}\qty{0.5}{\milli \meter}$).}
  \label{fig:trap_axiaView_zoomed_in}
\end{figure}

\begin{figure}[ht!] 
  \centering
  \includegraphics[width=0.75\columnwidth]{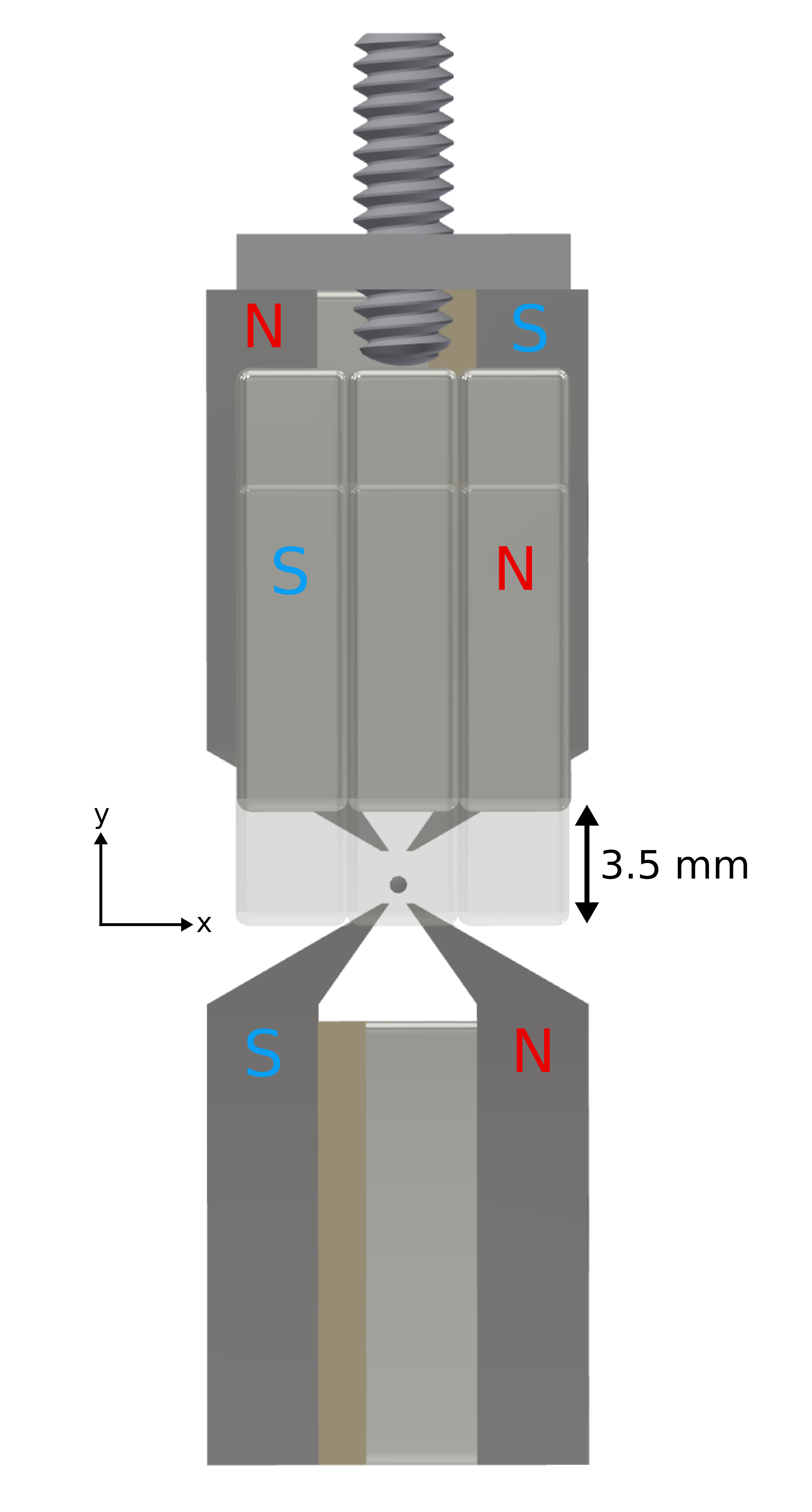}
  \caption{A 3-D scaled drawing displaying the total usable range of the end-magnet stacks and the direction of their magnetization relative to the magnets between the pole pieces.}
  \label{fig:trap_magnets_upAnddown}
\end{figure}

\begin{figure*}[ht!] 
  \centering
  \includegraphics[width=\textwidth]{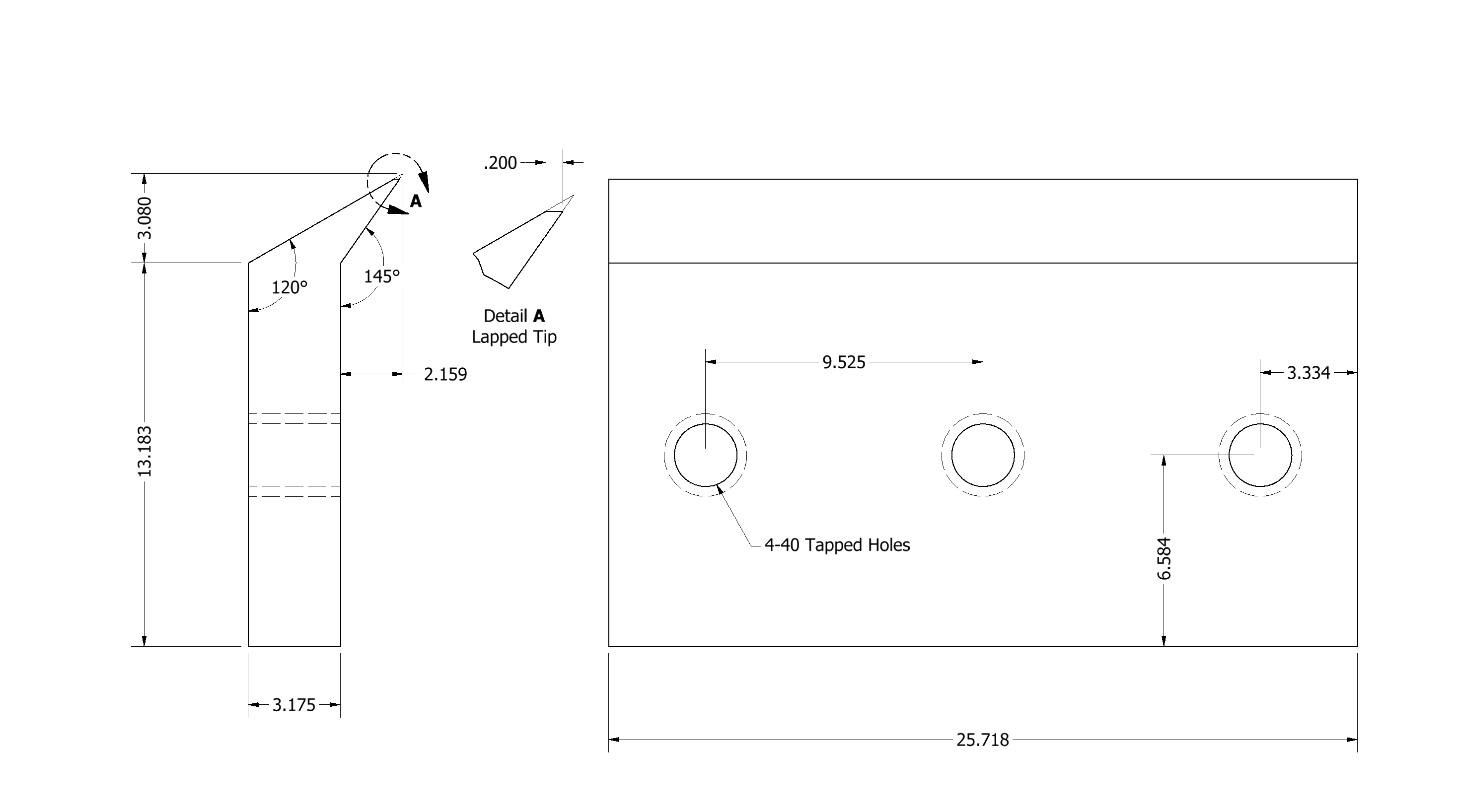}
  \caption{A 2-D drawing showing the dimensions of the bottom pole pieces used in the trap. All units are in millimeters.}
  \label{fig:bottom_polePiece_schematic} 
\end{figure*}

\begin{figure*}[ht!] 
  \centering
  \includegraphics[width=\textwidth]{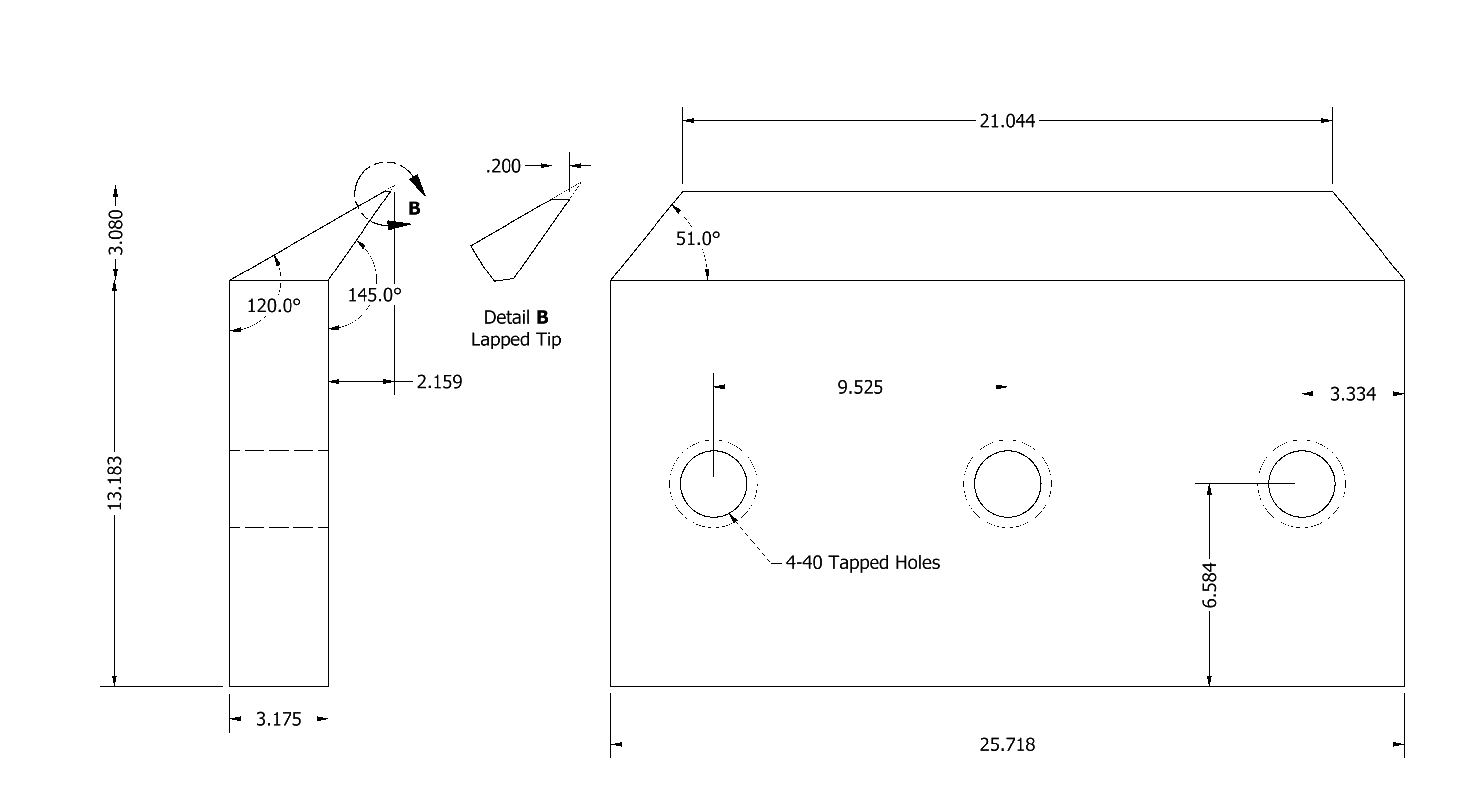}
  \caption{A 2-D drawing showing the dimensions of the top pole pieces used in the trap. All units are in millimeters.}
  \label{fig:top_polePiece_schematic}
\end{figure*}

\begin{figure*}[ht!] 
  \centering
  \includegraphics[width=\textwidth]{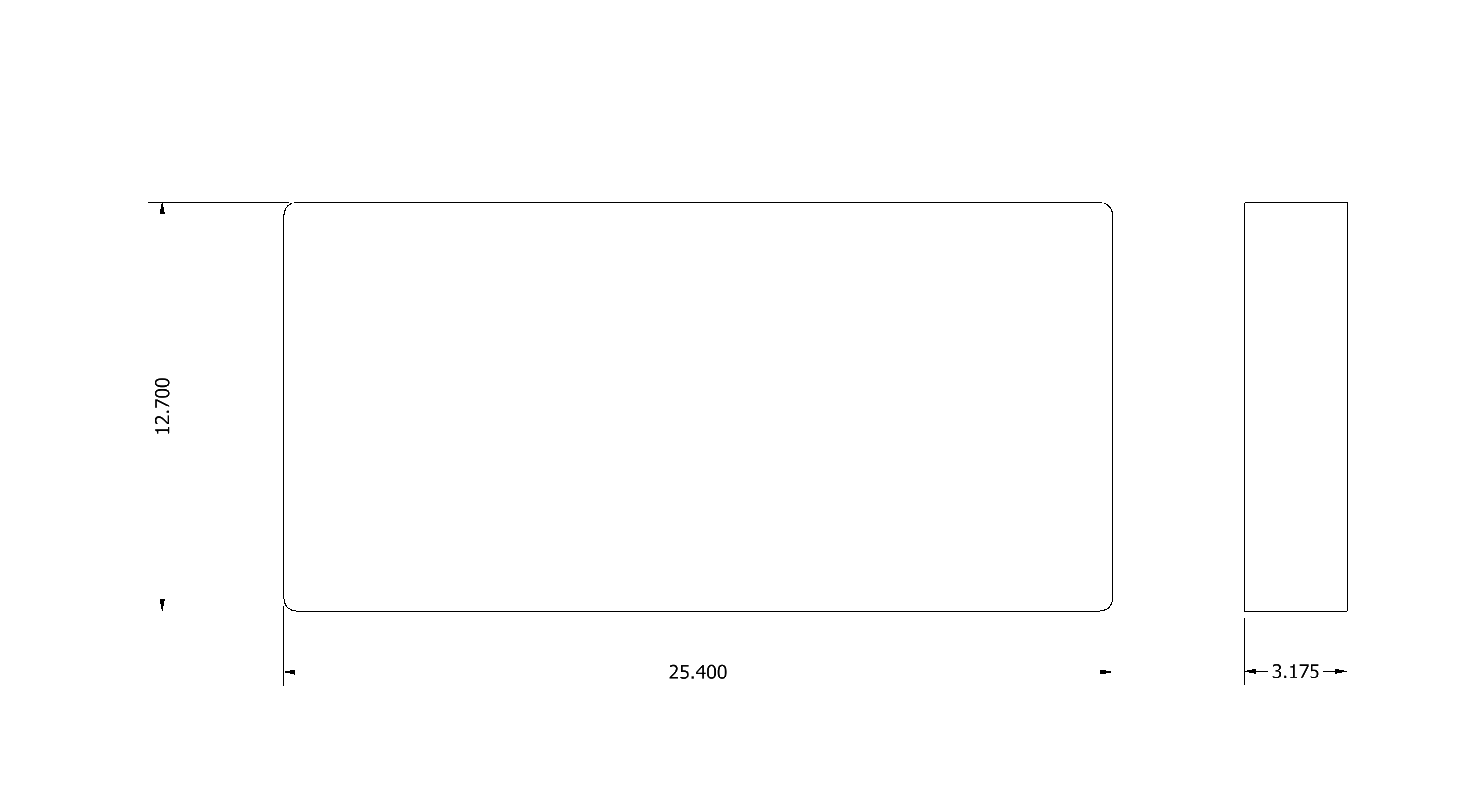}
  \caption{A 2-D drawing showing the dimensions of the NdFeB magnets used in the trap between the top and bottom pole pieces. All units are in millimeters.}
  \label{fig:magnet_schematic}
\end{figure*}

The following descriptions and figures provide further details of the magneto-gravitational trap design used in this experiment (\Cref{fig:trap_axiaView_zoomed_in,fig:trap_magnets_upAnddown,fig:bottom_polePiece_schematic,fig:top_polePiece_schematic,fig:magnet_schematic}).

The trap is composed of two nickel-plated NdFeB magnets (grade N42SH) (\Cref{fig:magnet_schematic}) surrounded by four Hiperco~50A (iron-cobalt alloy) pole pieces (\Cref{fig:bottom_polePiece_schematic,fig:top_polePiece_schematic}). Between the magnets and pole pieces are shims composed of a combination of steel, copper, polyimide film, and PEEK sheets of various thicknesses. The shim's thickness varies based on the desired transverse gap between the tips of the pole pieces. The magnets between the pole pieces are oriented with their magnetizations opposite of each other to produce a linear quadrupole field in the $xy$-plane. The combined asymmetry of the geometry of the pole pieces in the vertical direction and the end-magnet stacks on the sides of the trap (see Fig.~1 in the main text) constrain the rod in the axial direction. The end-magnet stacks are composed of three nickel-plated NdFeB (grade N42SH) magnets that are \qty{0.5}{"} $\times$ \qty{0.25}{"} $\times$ \qty{0.125}{"} thick. All the magnets used in the trap are sourced from K\&J Magnetics, Inc:
\begin{itemize}
    \item End-magnet stacks - Part No. B842SH
    \item Pole-piece magnets - Part No. BX082SH
\end{itemize}

As mentioned briefly in the main text, the adjustability of the side magnet stacks provide the possibility to tune the axial frequency of the rod's motion. That adjustability is possible through the use of adjustment screws that push down on the magnets (\Cref{fig:trap_magnets_upAnddown}) over an usable range of about \qty{3.5}{\milli \meter}. For this paper the magnets rested near the bottom of that range.  

\clearpage
\bibliography{bibliography_capfixed}